\documentstyle[preprint,aps]{revtex}
\begin{document}

\title{Differential cross sections for elastic
and inelastic positronium-hydrogen-atom scattering}

\author{Sadhan K. Adhikari}
\address{Instituto de F\'{\i}sica Te\'orica, 
Universidade Estadual Paulista
01.405-900 S\~ao Paulo, S\~ao Paulo, Brazil\\}

\date{\today}
\maketitle

\begin{abstract}

We report results of differential cross sections for elastic scattering,
target-elastic Ps excitations and target-inelastic
 excitation of hydrogen in a five-state coupled-channel model allowing for
Ps(1s)H(2s,2p) and Ps(2s,2p)H(1s)  excitations using a recently proposed
time-reversal-symmetric regularized electron-exchange model potential.
The present model yields a singlet Ps-H S-wave resonance at 4.01 eV of
width 0.15 eV and a P-wave resonance at 5.08 eV of width 0.004 eV. 
 We
also study the effect of the inclusion of the excited Ps and H states on
the convergence of the coupled-channel scheme.

{\bf PACS Number(s):  34.10.+x, 36.10.Dr}

\end{abstract}


\newpage

Lately, there have been great interest in the experimental \cite{1,2,3,3a}
and theoretical \cite{4,5,5x} studies of ortho positronium (Ps)  atom
scattering from different neutral atomic and molecular targets.  We
suggested a time-reversal symmetric regularized nonlocal electron-exchange
model 
potential \cite{6} and used it in the study of total cross section of Ps
scattering
by H \cite{7}, He \cite{6,8}, Li \cite{5t}, Ne \cite{9}, Ar \cite{9} and
H$_2$
\cite{10}. Our results were in agreement with experimental total cross
section \cite{1,2}, specially at low energies for He, Ne, Ar and H$_2$.
Among all Ps-atom systems, the positronium-hydrogen (Ps-H) system is the
simplest and is of fundamental interest.

In our first studies on Ps-H system we calculated the partial cross
sections, low-energy phase shifts, scattering lengths and effective ranges
and S-wave singlet resonance and binding energies using a regularized
model-exchange potential \cite{7,8}. This potential has a parameter which
allows for small variations of the scattering observables at low energies.
The results for scattering at high energies is insensitive to the
variation of
this parameter. At medium and high energies the cross sections of the
present model reduce to \cite{am} the first Born cross sections with
Oppenheimer
exchange potential \cite{14}.  The present model is made to 
reproduce simultaneously the accurate variational estimates for 
 the singlet S-wave resonance \cite{11} and Ps-H binding
energies
\cite{12}.  The agreement of the calculated singlet resonance and binding
energies with the variational estimates assures of the realistic nature of
the regularized exchange potential \cite{7}. Hence, for the sake of
completeness a study of the differential cross sections for Ps-H
scattering with the regularized electron-exchange potential seems
worthwhile.

In this paper we present a theoretical study of Ps-H scattering
 employing a five-state model allowing for excitation of
both Ps and H atoms using the regularized model exchange potential
mentioned above.
We calculate the differential cross
sections for quenching in addition to those 
for different elastic and inelastic transitions. We also report the
energies and widths of the singlet Ps-H resonances in different partial
waves.

The theory for the coupled-channel study of Ps-H scattering with the
regularized model potential has already appeared in the literature
\cite{6,7,8}. It is worthwhile to quote the relevant working equations
here.  We solve the following Lippmann-Schwinger scattering integral
equation in momentum space \begin{eqnarray} f^\pm_{\mu '\nu ',\mu\nu} (
{\bf k',k})&=& {\cal B}^\pm _{\mu '\nu ',\mu\nu}({\bf k ',k})  \nonumber
\\ &-&\sum_{\mu ",\nu "} \int \frac{ {\makebox{d}{\bf k"}} } {2\pi^2}
\frac { {\cal B}^ \pm _ {\mu '\nu ',\mu" \nu"} ({\bf k ',k"}) f^ \pm
_{\mu" \nu" ,\mu\nu} ({\bf k",k}) } {{k}^2_{\mu "\nu "}/4-k"^2/4+
\makebox{i}0} \label{4} \end{eqnarray} where the singlet (+) and triplet
($-$) ``Born"
amplitudes, ${\cal B}^\pm$, are given by $
 {\cal B}^\pm_{\mu '\nu ',\mu\nu}({\bf k',k}) = 
 g^D_{\mu '\nu ',\mu\nu}({\bf
k',k})\pm g^ E_{\mu '\nu ',\mu\nu}({\bf k',k}),$ 
   where $g^D$ and $g^E$ represent the direct and exchange Born amplitudes
and the $f^ \pm$ are the singlet and triplet scattering amplitudes,
respectively. The quantum states are labeled by the indices $\mu\nu$,
$\mu$ referring to the hydrogen atom and $\nu$ to the Ps atom. The
variables ${\bf k}$, ${\bf k'}$, ${\bf k''}$ etc denote the appropriate
momentum states; ${\bf k}_{\mu ''\nu ''}$ is the on-shell relative
momentum of Ps with respect to H in the channel $\mu ''\nu ''$.  We use
units $\hbar = m = 1$ where $m$ is the electron mass. 
  The differential cross section is defined 
by 
\begin{equation}
\left(\frac{d\sigma}{d\Omega}\right)_{\mu '\nu ',\mu\nu} = \frac
{k'}{4k }
\left[|f^ +_{\mu '\nu ',\mu\nu}({\bf k',k})|^2+
3|f^ -_{\mu '\nu ',\mu\nu}({\bf k',k})|^2   \right]
\end{equation}
and the quenching cross section that describes conversion from ortho- to 
para-positronium is defined
by 
\begin{equation}\left(\frac{d\sigma}{d\Omega}\right)_{\mu '\nu ',\mu\nu}^
{\makebox{quen}} = \frac {k'} {16k}
\left|f^+_
{\mu '\nu ',\mu\nu}({\bf k',k}) -f^- _{\mu '
\nu ',\mu\nu}({\bf k',k})\right|^2.
\end{equation}
The Ps-H direct Born  amplitude  is given by \cite{15}
\begin{eqnarray}\label{1x} 
g^D_{\mu'\nu',\mu\nu} ({\bf k_f,k_i})&= &
\frac{4}{Q^2}
\int \phi^*_{\mu '}({\bf r})\left[ 1-\exp ( \makebox{i} {\bf Q.
r})\right]\phi_\mu({\bf r})
\makebox{d}{\bf r}\nonumber \\ &\times&
\int \chi^*_{\nu '}({\bf  t })2\mbox{i} \sin ( {\bf Q}.{\bf  t
}/
2)\chi_{\nu}({\bf  t } ) \makebox{d}{\bf  t }.
\end{eqnarray}
The Ps-H  model exchange (Born) amplitude  is a generalization of the
electron-hydrogen model exchange potential of ref. \cite{r} and 
is given by \cite{8}
\begin{eqnarray}\label{1} 
g^E_{\mu'\nu',\mu\nu} ({\bf k_f,k_i})&= &
\frac{4(-1)^{l+l'}}{D}
\int \phi^*_{\mu '}({\bf r})\exp ( \makebox{i} {\bf Q. r})\phi_\mu({\bf r})
\makebox{d}{\bf r}\nonumber \\ &\times&
\int \chi^*_{\nu '}({\bf  t })\exp ( \makebox{i}{\bf Q}.{\bf  t }/
2)\chi_{\nu}({\bf  t } ) \makebox{d}{\bf  t }
\end{eqnarray}
with 
\begin{equation}\label{2}
D=(k_i^2+k_f^2)/8+C^2[(\alpha_\mu^2+\alpha_{\mu '}^2)/2+(\beta_\nu^2+
\beta_{\nu'}^2)/2]
\end{equation}
where $l$  and $l'$ are  the angular momenta of the initial and final Ps
states, 
the initial and 
final Ps momenta are ${\bf k_i}$  and ${\bf k_f}$, ${\bf Q = k_i -k_f}$,
$\alpha_\mu^2/2$ and $\alpha_{\mu ' }^2/2$, and $\beta_\nu^2$
and $\beta_{\nu '}^2$ are the binding energies of the initial and final
states of  H and  Ps in atomic units, 
respectively, and $C$ is the only parameter of the potential. Normally,
the parameter $C$ is taken to be unity which leads to reasonably good
result. However, it can be varied slightly from unity to get a precise fit
to a low-energy observable. This variation  of $C$ has no effect on the
scattering observables at high energies and the model exchange potential
reduces to \cite{am} the Born-Oppenheimer exchange potential \cite{14} at
high
energies.

After a partial-wave projection, the singlet (+) and triplet ($-$) 
scattering equations  (\ref{4})  are solved by the
method of matrix inversion. The maximum number of partial waves included
in the
solution of the integral equation  is 100 and the contribution of
the higher partial waves is included in the first Born approximation.  
This procedure  provides convergence of the partial-wave
scheme.

In the present study we use
the value $C=0.785$ throughout the
present investigation,  as  in a recent  study 
\cite{8}.  
Interestingly enough, with this value of $C$, the
 five-state model produces a singlet S-wave Ps-H binding energy of 1.05
eV and resonance energy of 4.01 eV with width 0.15 eV. This binding
energy is consistent
with both the accurate 
variational  estimate of 1.067 eV
\cite{12} and experimental result of $1.1\pm0.2$ eV \cite{13}. Whereas
the present resonance energy is essentially identical to the
recent variational
study  of 4.0058 eV, the agreement of  the present width of 0.15 
eV 
 with the  variational estimate of 0.0952 eV is  only fair \cite{11}.
The 22-Ps-state R-matrix calculation \cite{4} yielded a S-wave resonance
energy of
4.55 eV with width 0.084 eV.
 The present model with $C=0.785$  yielded a P-wave
resonance at 5.08 eV with width 0.004 eV. Variational calculation for the 
P-wave resonance \cite{13ho} yielded a energy of 4.285 eV with
width 0.0435
eV, whereas the 22-Ps-state
R-matrix calculation  yielded 4.88 eV and 0.058 eV for these
quantities, respectively.

Here we present results of Ps-H scattering using a five-state model 
that includes the 
 following states: Ps(1s)H(1s), Ps(2s)H(1s),
Ps(2p)H(1s), Ps(1s)H(2s) and Ps(1s)H(2p).  
The truncated model  that includes the
first
$n$ states of this set will be termed the $n$-state model.  The Born terms
for the  simultaneous excitation of both H and Ps atoms are found to be 
 small and will not be
considered here in the coupled-channel scheme.  
We calculate the elastic
Ps(1s)H(1s)  differential cross section and inelastic differential  cross
sections to
Ps(2s)H(1s), Ps(2p)H(1s), Ps(1s)H(2s) and Ps(1s)H(2p) states. In addition
we calculate the differential quenching cross section for elastic
scattering. 

In order to show the general trend of the result, we performed
calculations at the following incident positronium energies: 
20, 30, 40, 60, and 100 eV. 
We exhibit the differential cross sections for elastic scattering and
differential quenching cross sections for elastic  scattering at these
energies in Figs. 1 and 2, respectively.
In Figs. 3 $-$ 6 we exhibit the inelastic cross sections for transition to 
Ps(2s)H(1s), Ps(2p)H(1s), Ps(1s)H(2s) and Ps(1s)H(2p) states
at 20, 30, 40, 60, and 100 eV. 
From all these figures we find that,  as expected, the differential cross
section is more isotropic at low energies where only the low partial waves
contribute to the cross sections. At higher energies more and more partial
waves are needed to achieve convergence and the differential cross
sections
are more anisotropic.

As Ps is highly polarizable compared to H, Ps-excitation cross sections
are larger than corresponding H-excitation cross sections and 
Ps excitations are expected to
play a more important role in Ps-H scattering compared to H excitations.
However, it is interesting to investigate the effect of the inclusion of H
and Ps 
states on the dominating inelastic Ps cross sections at low and medium
energies. At
high energies such effect is small and the different cross sections tend
to  the corresponding first Born approximation cross sections with the 
the Oppenheimer exchange potential \cite{am,14}. 
Hence,  in addition to just reporting the differential cross sections we
also
study the effect of including more states in the expansion scheme to the
dominating Ps(1s,2s,2p)
partial cross sections. We make this study on the partial  cross
sections  
at different energies which we show in
Table I. From the table we see how  the elastic cross
sections change with the inclusion of more Ps functions in the
basis set. 
However, the inclusion of basis functions corresponding to
higher excitations of H states have less  influence on the
convergence.  Such small change of   partial cross sections
as 
in Table I will contribute to a small difference in the corresponding
differential  cross
sections which will be unobservable in a plot on a logarithmic scale as
 in Figs. 1 $-$ 6. Hence we do not exhibit the corresponding  differential
cross
sections here.

\vskip .2cm {Table 1: Ps-H partial cross sections in units of $\pi
a_0^2$ at different positronium energies using 
different basis functions} \vskip .2cm
\begin{centering}

\begin{tabular} {|c|c c c c c|c c c c|c c c|}
\hline 
E & Elast&Elast & Elast&Elast  &Elast&Ps(2s) &Ps(2s)&Ps(2s)&Ps(2s)&Ps(2p)
&
Ps(2p) & Ps(2p) \\
eV & 1-St & 2-St &3-St & 4-St &5-St&2-St&3-St&4-St&5-St &3-St&4-St &5-St\\
  \hline
 0.1&24.18 &22.91  & 21.52  &20.73 & 19.84       &  & & & &     & & \\
5&  7.28 & 7.24  &     5.51  & 5.47& 5.29    &       & & & & & & \\
10&3.97 &3.74 &    2.80        & 2.90&2.50& 0.117 &0.115&0.118&0.120 &
1.68
&1.74 & 1.72\\
 20& 1.41& 1.37 &1.20            & 1.18     & 1.18& 0.078&0.074&0.070&
0.069&
0.83
& 0.83 & 0.82\\
\hline
\end{tabular}

\end{centering}

\vskip 0.2cm

To summarize, we have performed a five-state coupled-channel calculation
of Ps-H scattering at medium energies using a regularized
time-reversal symmetric
electron-exchange model 
potential recently suggested by us and successfully used in other Ps
scattering problems. The present model yields a singlet Ps-H S-wave 
resonance at 4.01 eV of
width 0.15 eV and a P-wave resonance at 5.08 eV of width 0.004 eV. 
We present results for differential cross sections at
several incident Ps energies between 20 eV to 100 eV for elastic
scattering. Differential cross sections of quenching scattering are also
reported in addition to those for elastic and inelastic   excitation to
Ps(2s,2p)H(1s) and Ps(1s)H(2s,2p) states. The effect of including the H
states in the coupled-channel scheme on the elastic and Ps-excitation
cross sections
is found to be small.

The work is supported in part by the Conselho Nacional de Desenvolvimento -
Cient\'\i fico e Tecnol\'ogico,  Funda\c c\~ao de Amparo
\`a Pesquisa do Estado de S\~ao Paulo,  and Finan\-ciadora de Estu\-dos e
Projetos of Brazil.

\newpage
{\bf Figure Caption:}

1. Differential cross section (in units of $a_0^2$) for elastic Ps-H
scattering at the following incident Ps energies: 20 eV (dashed-dotted
line), 30 eV (dashed-double-dotted line), 40 eV (dashed-triple-dotted
line), 60 (dashed line), and  100 eV (full line). 

2. Differential quenching cross section (in units of $a_0^2$) for elastic
Ps-H
scattering at the following incident Ps energies: 20 eV (dashed-dotted
line), 30 eV (dashed-double-dotted line), 40 eV (dashed-triple-dotted
line), 60 (dashed line), and  100 eV (full line). 

3. Differential cross section (in units of $a_0^2$) for inelastic
Ps-H
scattering to Ps(2s)H(1s) state at the following incident Ps energies: 20
eV (dashed-dotted
line), 30 eV (dashed-double-dotted line), 40 eV (dashed-triple-dotted
line), 60 (dashed line), and  100 eV (full line). 

4. Differential cross section (in units of $a_0^2$) for inelastic
Ps-H
scattering to Ps(2p)H(1s) state at the following incident Ps energies: 20
eV (dashed-dotted
line), 30 eV (dashed-double-dotted line), 40 eV (dashed-triple-dotted
line), 60 (dashed line), and  100 eV (full line).

5. Differential cross section (in units of $a_0^2$) for inelastic
Ps-H
scattering to Ps(1s)H(2s) state at the following incident Ps energies: 20
eV (dashed-dotted
line), 30 eV (dashed-double-dotted line), 40 eV (dashed-triple-dotted
line), 60 (dashed line), and  100 eV (full line). 

6. Differential cross section (in units of $a_0^2$) for inelastic
Ps-H
scattering to Ps(1s)H(2p) state at the following incident Ps energies: 20
eV (dashed-dotted
line), 30 eV (dashed-double-dotted line), 40 eV (dashed-triple-dotted
line), 60 (dashed line), and  100 eV (full line). 

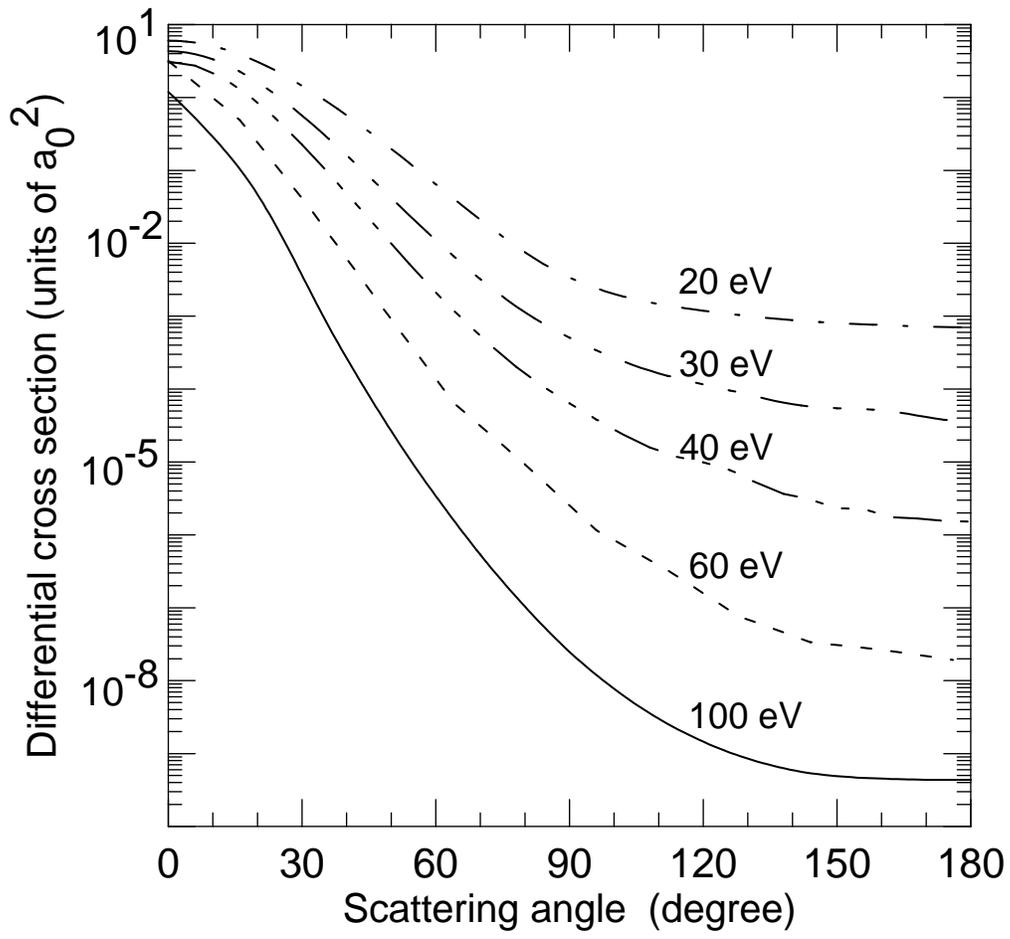
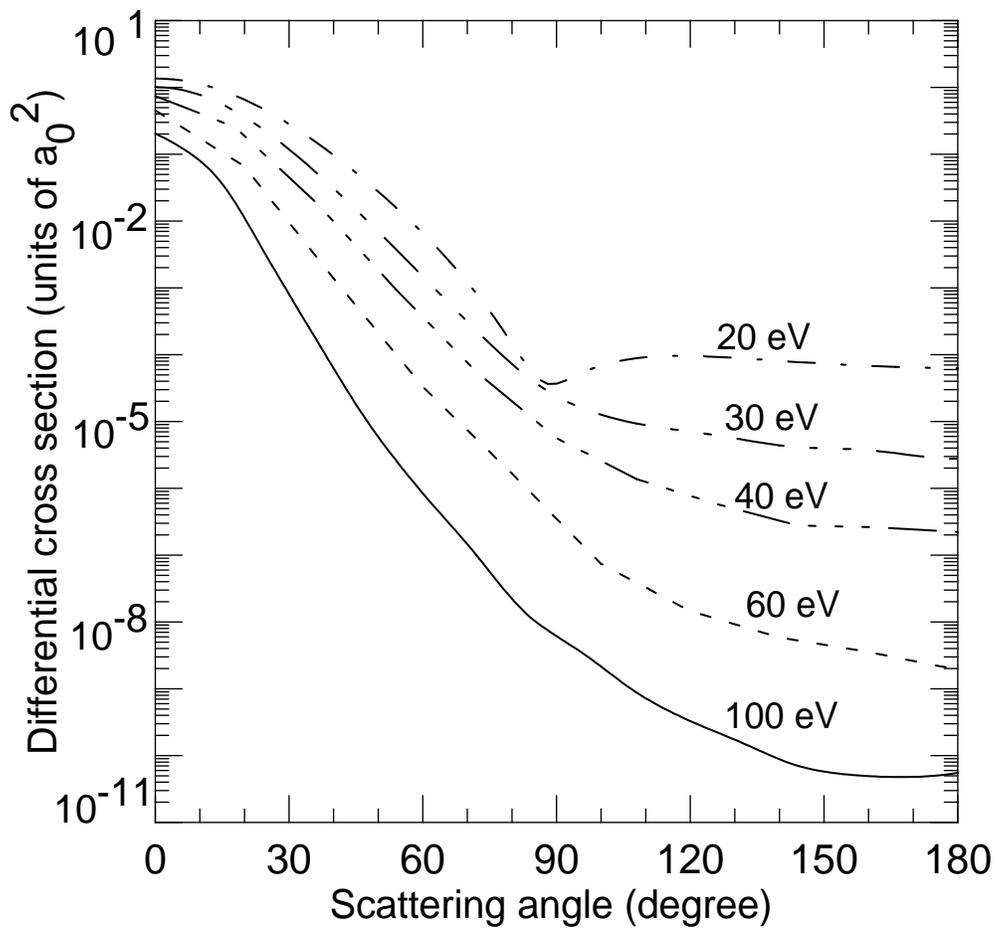
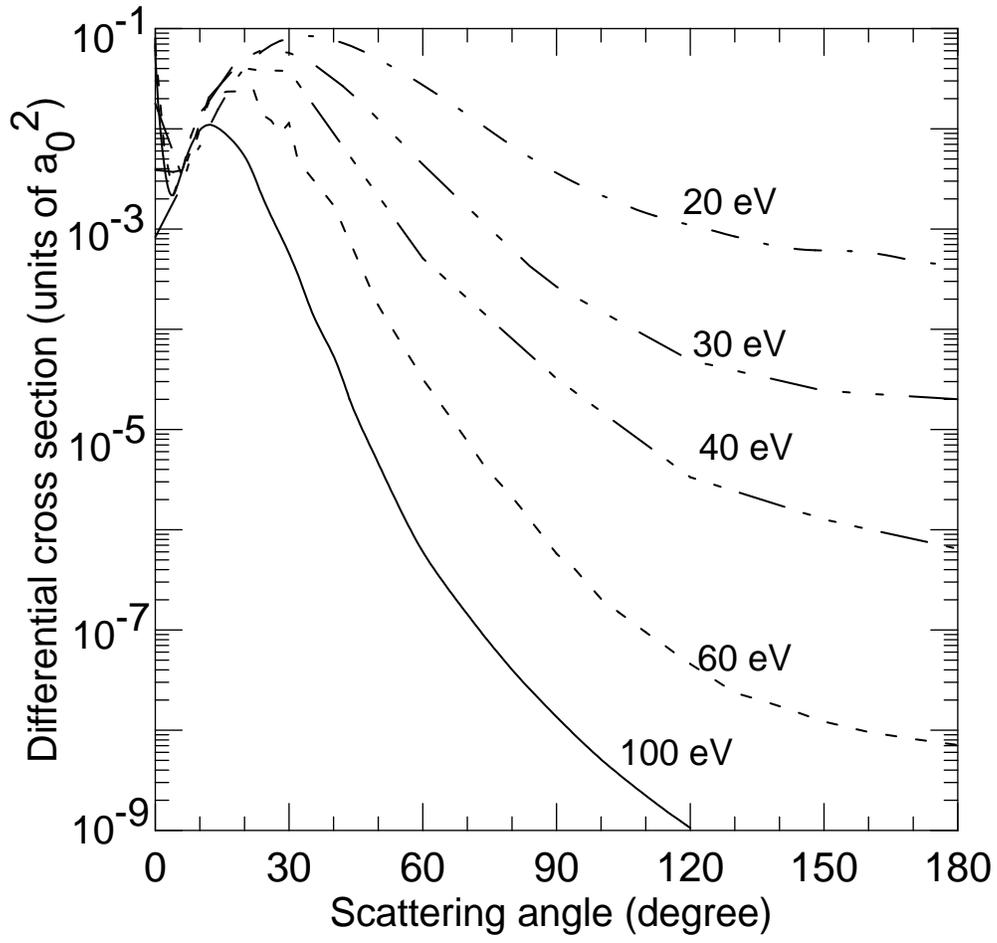
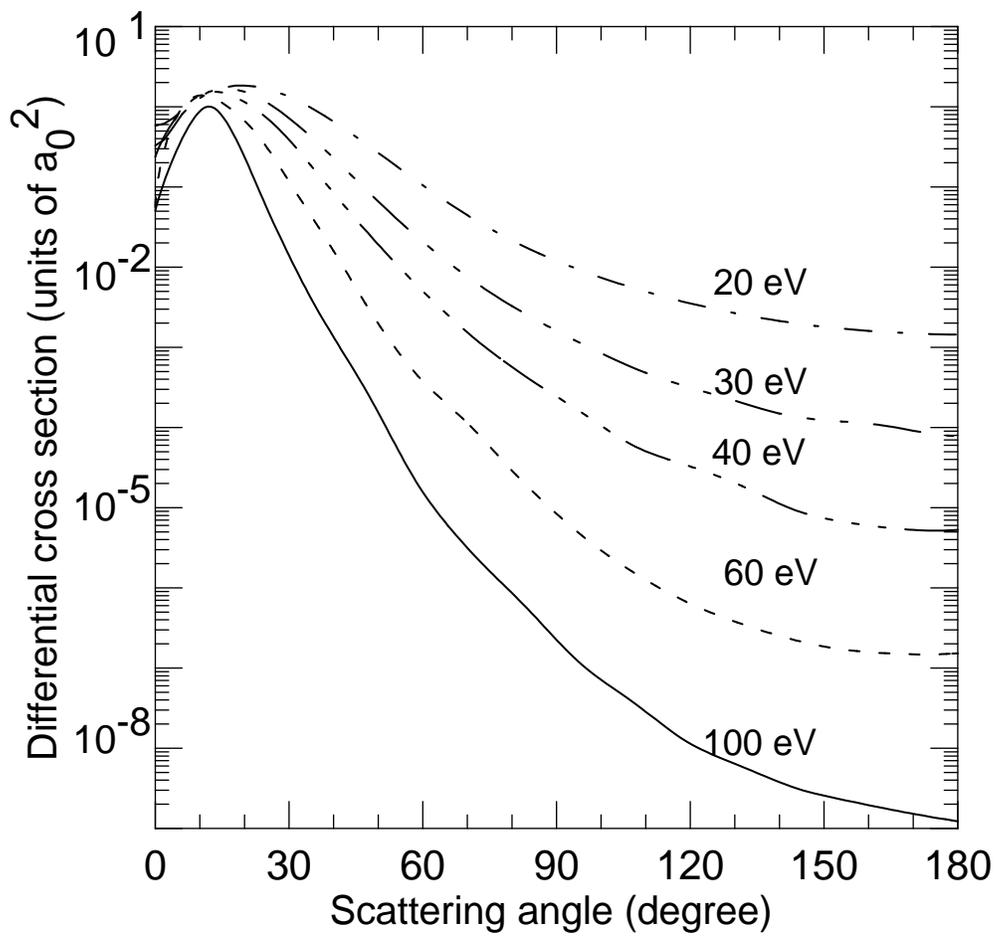
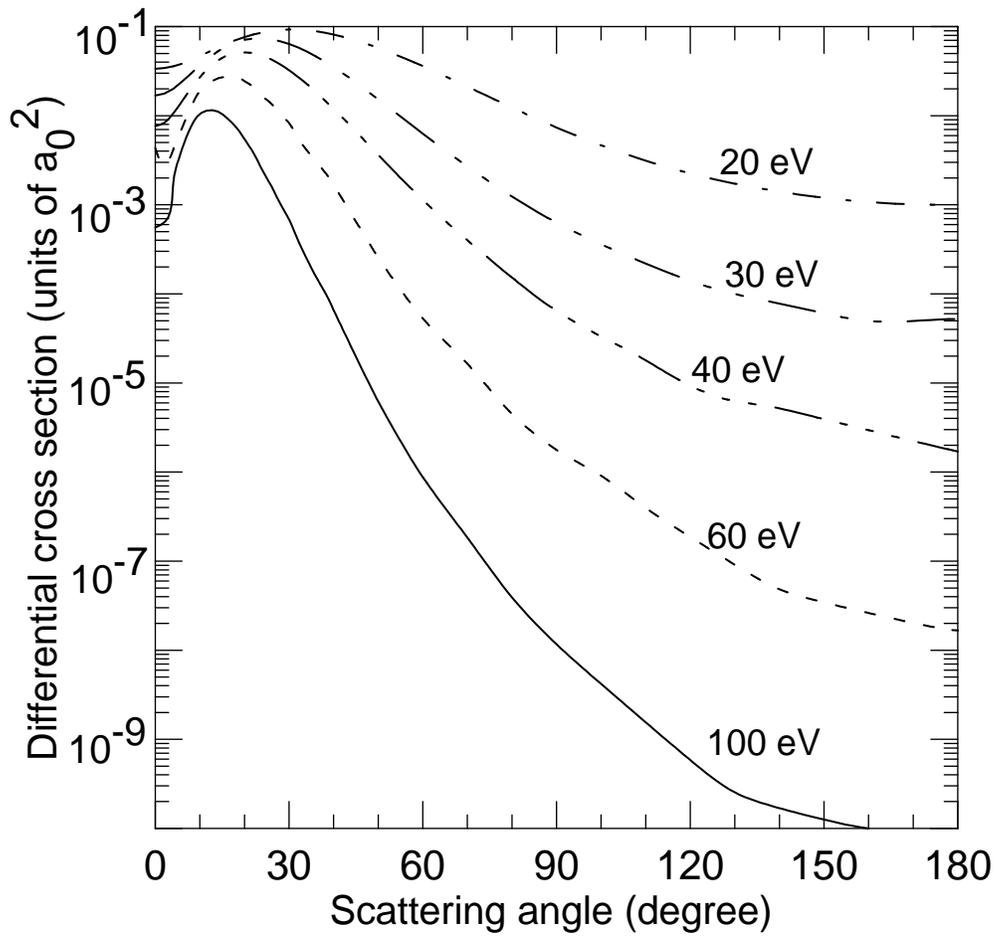
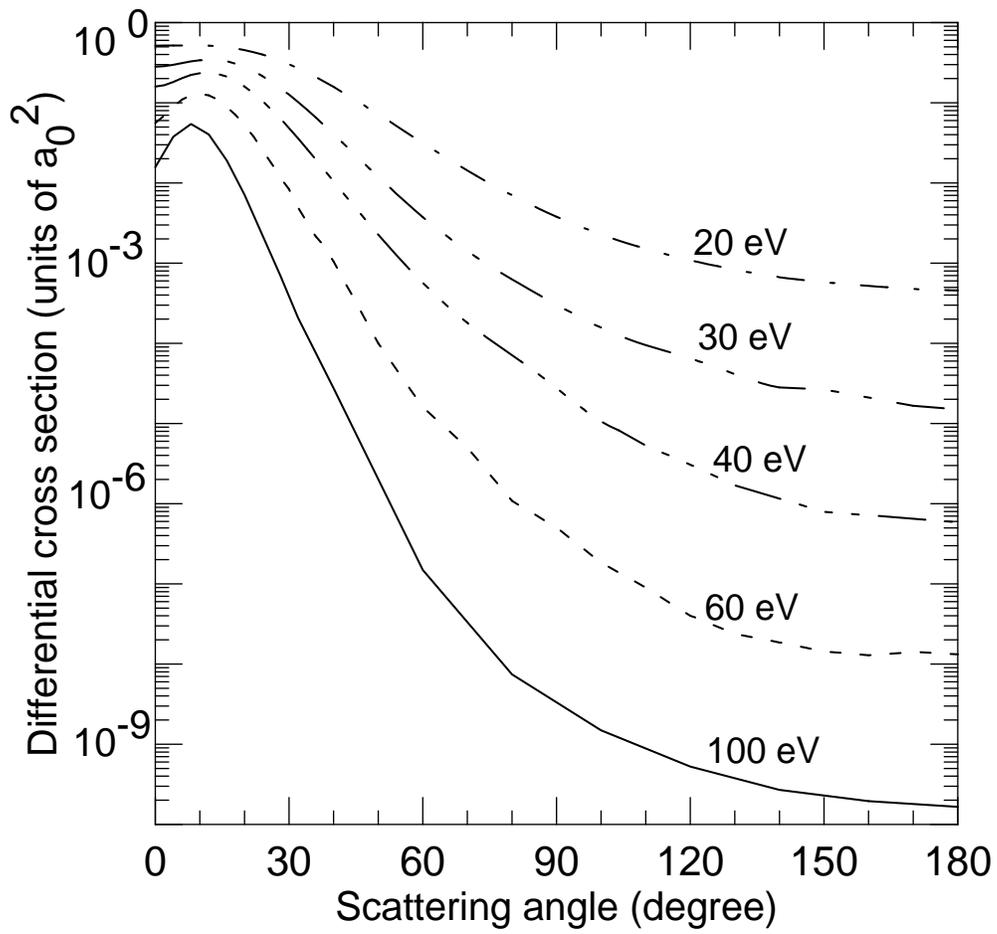
\end{document}